\documentclass{article}
\usepackage{graphicx}
\begin{document}

\title{ASTRONOMY IN ARGENTINA}
\author{HERN\'AN MURIEL\\
Observatorio Astron\'omico de C\'ordoba\\
and IATE (CONICET-UNC)\\
Laprida 854. X5000 BGR. C\'ordoba, Argentina\\
\texttt{hernan@oac.uncor.edu}}
\maketitle

\begin{abstract}
This article analyses the current state of Astronomy in Argentina and 
describes its origins. We\footnote{ Hern\'an Muriel was President of the Argentinian 
Astronomical Society for the period 2008-2011.} briefly describe the institutions where
astronomical research takes place, the observational facilities available, the
training of staff and professionals, and the role of the institutions in scientific 
promotion. We also discuss the outreach of Astronomy towards the
general public, as well as amateur activities. The article ends with an analysis of 
the future prospects of astronomy in Argentina\footnote{Although we have tried to be 
as objective as possible, some statements inevitably contain some personal views.}.
\end{abstract}

\section{Introduction}

Argentina has an important scientific tradition, in which astronomy has
been one of the pioneering areas. There are interesting documents recording
astronomical observations carried out at the beginning of the 18th century,
but systematic studies of astronomy in Argentina began with the creation
of the Observatorio Nacional Argentino (ONA, National Argentine Observatory, 
today the Observatorio Astron\'omico de C\'ordoba, see Fig. 1) in
C\'ordoba on 14 Oct 1871. The President, Domingo Faustino Sarmiento, entrusted the 
study of the southern skies to the renowned US astronomer
Benjamin Gould (who was the founder, among other projects, of The Astronomical 
Journal). With ONA's foundation, Sarmiento's intention was
to send a clear message to the country and to the world, as he declared
on the occasion of the inauguration of the new observatory: {\it "We should
renounce our standing as a Nation, or our title as a civilized people, if we
do not take our part in the progress and development of natural sciences"}.
Some historians assign such importance to ONA's creation, and particularly to 
Benjamin Gould's presence, that they consider that moment as the
introduction of modern science in Argentina.

%--------------- 2

%------------------------- FIG1
\begin{figure}[h!]
  \caption{The Astronomical Observatory of C\'ordoba. (\copyright Hern\'an Muriel)}
  \centering
    \includegraphics[width=0.9\textwidth]{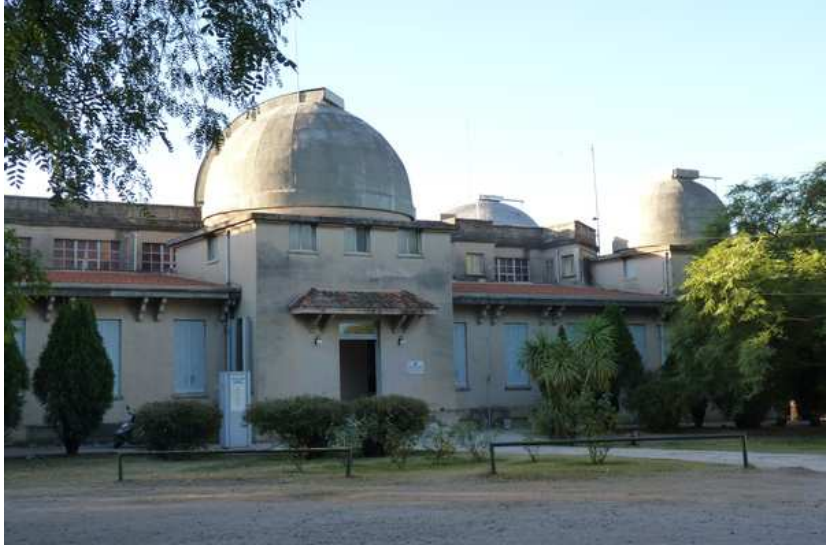}
\end{figure}

The second milestone in the history of astronomy in Argentina was the
creation of the Astronomical Observatory of La Plata in 1883 (see Fig. 2).
This institution was the cradle of the first school of astronomy, founded in
1935, where the first Argentinian astronomers were formed, including local
pioneer astronomers such as Jorge Sahade, Jos\'e Luis S\'ersic and Carlos Cesco.

In spite of these strong beginnings, the growth of science in Argentina
has not been easy, as it has been always complicated by the to-and-fros
of the country's politics, with periods of great impetus mixed with other,
very dark times, which resulted in a significant exodus of Argentinian scientists, 
who found positions in research centres all over the world. Amid
this constant flux, one undeniably positive fact has been the creation of the
{\it Consejo Nacional de Investigaciones Cient\'{\i}ficas y T\'ecnicas} 
(CONICET\footnote{http://www.conicet.gov.ar/},
National Council for Scientific and Technical Investigation) in 1958; nowadays, 
together with the universities, CONICET provides essential support
to the development of science and particularly to astronomy in this country.
Currently, one hundred and forty years after ONA's foundation, there are
a variety of Argentine astronomical institutions, where about two hundred
researchers work on a wide range of topics.

%------------------- 3

%------------------------- FIG2
\begin{figure}[h!]
  \caption{The Astronomical Observatory of La Plata. (courtesy Guillermo Sierra)}
  \centering
    \includegraphics[width=0.9\textwidth]{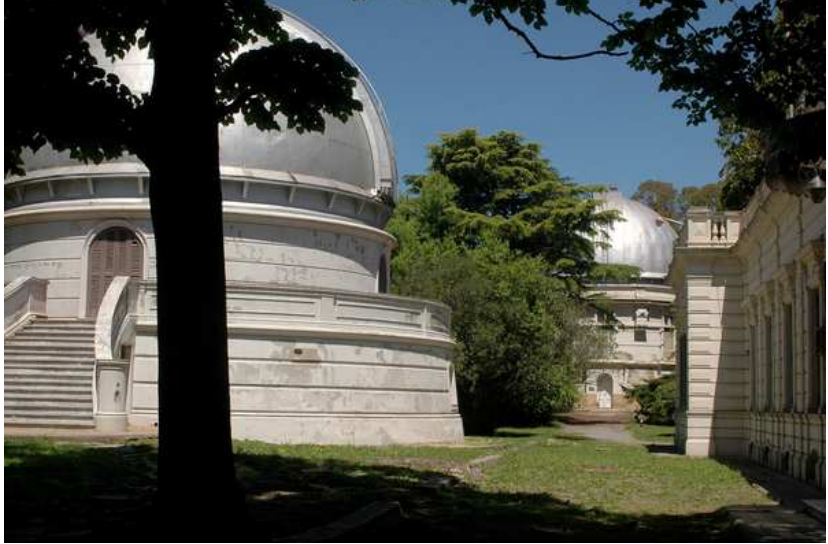}
\end{figure}

\section{Astronomical Institutions in Argentina}

There are approximately a dozen institutions where astronomical research is
taking place in Argentina. In spite of having started with a defined research
profile in one specific area, most of these nowadays have scientists doing
research in a great diversity of topics (see Fig. 3), including the
most conventional fields along with novel ones.
Since the second half of the last century, a series of research institutes depending 
from CONICET have been created, some of which function within
the observatories, and in most cases depend on the universities as well.
From a geographical point of view, the vast majority of astronomical research in 
Argentina is centred in three provinces: Buenos Aires, C\'ordoba
and San Juan, located in the east-centre, centre and west-centre of the
country, respectively (see Fig. 4); however, in the future, more groups may
well consolidate in other regions where incipient activity is occurring.

%------------------ 4

%------------------------- Table 1 
\begin{figure}[h!]
  \caption{Principal research areas per institution.}
  \centering
    \includegraphics[width=1.\textwidth]{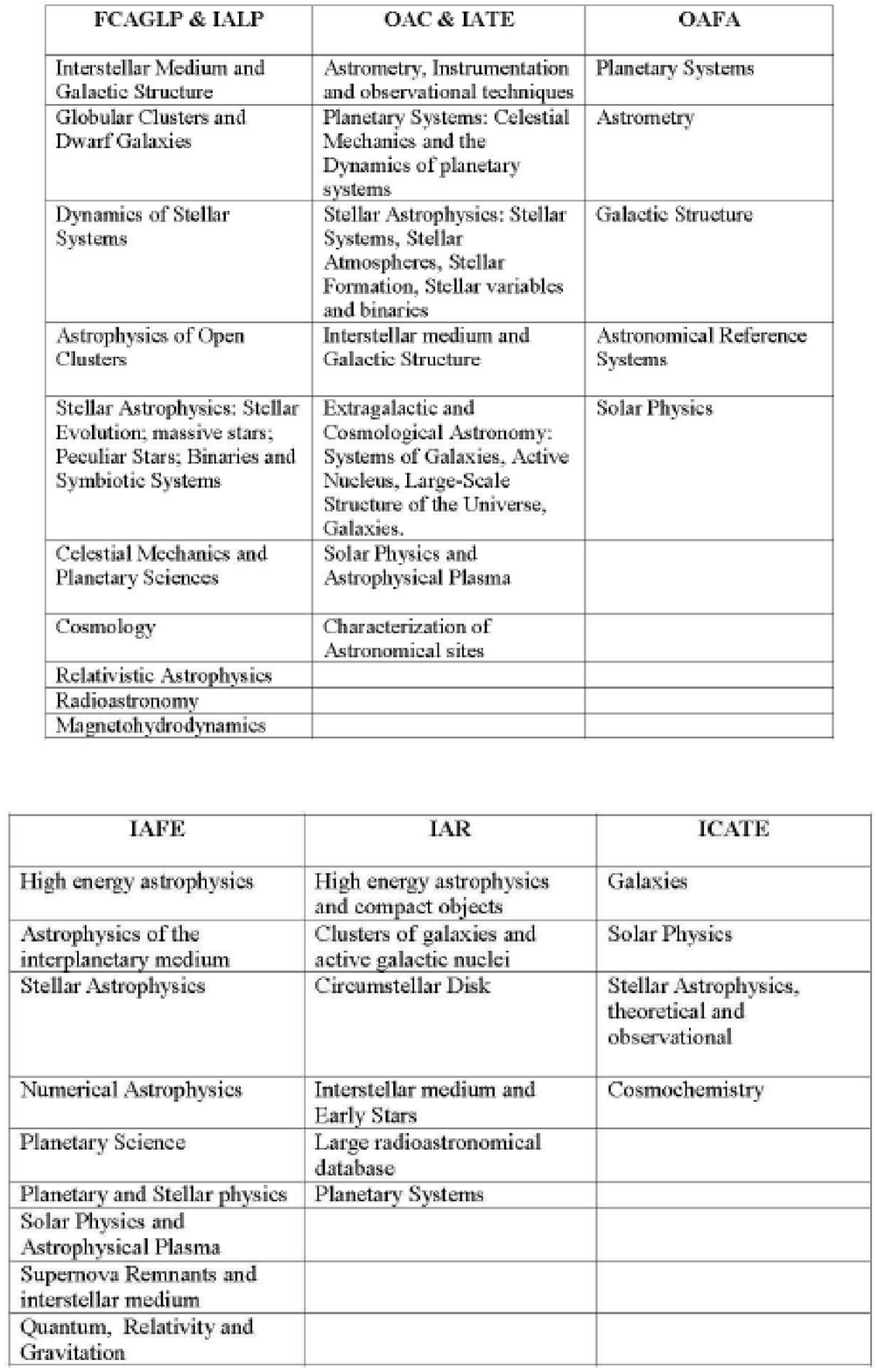}
\end{figure}

The following is a summary of institutions whose main activity is astronomical research:

%------------------- 5

%------------------------- Fig 3 
\begin{figure}[h!]
  \caption{A map of Argentina showing the location of observatories (in red), institutes of
CONICET where the main activity is astronomy (in green), institutes of CONICET where
astronomy is a secondary activity (in light blue), astronomical facilities in the mountains
(in yellow), planetariums (in black), characterized sites for astronomical facilities (in
orange), and Universities where Astronomy or related sciences are taught (in brown).
(graphics by Victor Renzi)}
  \centering
    \includegraphics[width=.9\textwidth]{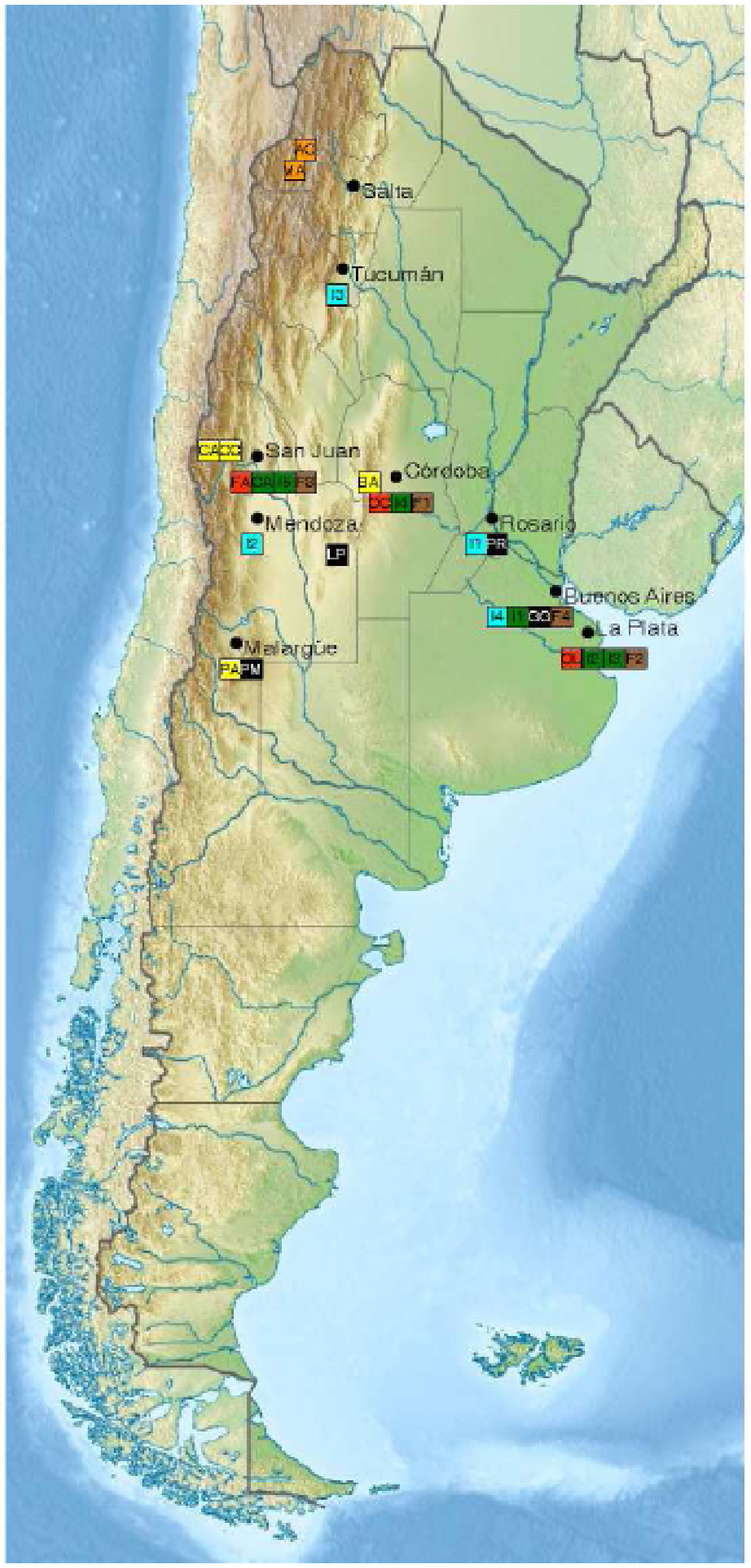}
\end{figure}

%------------------------ 6

\vskip .3cm
\noindent {\it Observatorio Astron\'omico de C\'ordoba (OAC) de la Universidad National
de C\'ordoba (UNC\footnote{Astronomical Observatory of the National University of C\'ordoba
(http://www.oac.uncor.edu/).})\\ 
\noindent Instituto de Astronom\'{\i}a Te\'orica y Experimental (IATE
\footnote{Institute for Theoretical and Experimental Astronomy (http://www.iate.oac.uncor.edu/).
}, C\'ordoba)}

As mentioned, astronomy in Argentina began with the creation of
the National Argentine Observatory (Observatorio Nacional Argentino) in
1871. After going through several changes in its institutional status, it finally 
became part of the National University of C\'ordoba (UNC), the oldest
university in the country (almost four hundred years old), and its second
largest.

The scientific profile of the institution was defined by the celebrated
Benjamin Gould, who initiated various astronomical projects, among which
the {\it Uranometry of C\'ordoba} was a pioneer work, cataloguing the position
and brightness of all stars visible to the naked eye from C\'ordoba. The
star survey {\it C\'ordoba Durchmusterung} was by no doubt one of the greatest
astronomical achievements in C\'ordoba during the nineteenth century. Argentina later 
also participated in the international {\it Carte du Ciel} project at
the beginning of the $20^{th}$ century.

Astrophysics arrived at the NAO with its director, Charles Perrine, who
also proposed and initiated the foundations for the construction of a big
reflector telescope. The project was completed with the help of Enrique
Gaviola, and inaugurated, in 1942, in the Astrophysical Station of Bosque
Alegre, 40km from the city of C\'ordoba. This 1.54m reflecting telescope
enabled C\'ordoba Observatory to position itself as a pioneer in astrophysical 
studies in the southern hemisphere, and to give a strong impulse to
extragalactic astronomy, led by Jos\'e Luis S\'ersic, who later published the
renowned {\it Atlas of Southern Galaxies\footnote{Published by the Astronomical Observatory of the UNC.
}} in 1968. In 1956, Gaviola initiated the
foundation of the Institute of Mathematics, Astronomy and Physics within
C\'ordoba National University, which in 1983 would become the Faculty of
Mathematics, Astronomy and Physics (FaMAF\footnote{http://www.famaf.unc.edu.ar/}), in which a considerable
number of astronomers have been formed.

The Astronomical Observatory of the UNC is today one of the most
important astronomical centres in the country, with approximately sixty
professors, who carry out research in varied areas of Astronomy. They are
also responsible for the teaching and training of graduate and postgraduate
astronomy students studying at the FaMAF.

%-------------------------------------------------- 7

Within the OAC, and under the auspices of the UNC and CONICET,
the Institute of Theoretical and Experimental Astronomy (IATE) was created, 
as a continuation of the Extra-Galactic Studies group founded by
S\'ersic more than three decades earlier. In spite of having started merely as
an extra-galactic research project, other areas have been incorporated since
then, such as solar physics, extra-solar planets and the search for new astronomical 
sites. The IATE nowadays has more than thirty post-graduate
researchers, many of whom are also professors in the OAC.

\vskip .3cm
\noindent {\it Observatorio Astron\'omico de La Plata (OALP\footnote{
Astronomical Observatory of La Plata, nowadays the Faculty of Astronomical Sciences 
and Geophysics of La Plata (Facultad de Ciencias Astron\'omicas y Geof\'{\i}ısicas de
La Plata, FCAGLP (http://www.fcaglp.edu.ar/).
})\\
\noindent Instituto de Astrof\'{\i}sica de La Plata (IALP\footnote{Institute of Astrophysics of La Plata
(http://www.fcaglp.unlp.edu.ar/∼gladys/ialp/).
})}

The creation of the Astronomical Observatory of La Plata was related
to Argentina's participation in the transit of Venus observations in 1882,
together with the need to endow the new city with public buildings. Among
the aims of founding the OALP was the realisation of mapping studies, for
which reason its first activities were basically of services. As from the thirties, 
systematic astronomical observations have been made of occultations,
eclipses, asteroids and comets.

It is important to note that the Observatory is older than the University 
of La Plata (UNLP), which, founded in 1905, is nowadays one of the
most important in the country. The insertion of the OALP into the UNLP,
together with F\'elix Aguilar's special interest, led to the creation of the first
School of Astronomy in the country in 1935, which was to have a strong
impact on the future of astronomy in Argentina. With its first graduates,
some of whom had completed further studies abroad, and the presence of
Livio Gratton, the Observatory of La Plata initiated its astrophysical studies, 
as well as the idea of developing a large telescope. This idea would be
materialized some decades later by Jorge Sahade, with the creation of the
Astronomical Complex of "Leoncito" (CASLEO) in the province of San
Juan.

In 1982, the OALP School of Astronomy became the Faculty of Astronomical 
Science and Geophysics (FCAGLP), permitting it to recover
from the harsh decade of the seventies and to begin a period of growth,
which has been sustained up to the present. The Observatory of La Plata
has traditionally offered the study of varied astronomical areas such as Astrometry 
and Stellar Astrophysics, though nowadays research is carried out
in most areas of contemporary astronomy. The FCAGLP today has more
than fifty professors in astronomical fields, and is one of the most important
astronomy institutions in Argentina

%------------------------------------ 8

In 1999, the Institute of Astrophysics of La Plata (IALP) was created,
housed in the FCAGLP and under the UNLP and the CONICET. In the
IALP, research is carried out in areas such as Dynamics, Stellar and Planetary 
Evolution and Formation. Most IALP researchers are also professors
at the FCAGLP, and its staff of more than fifty includes graduate students
and postdoctoral fellows.

\vskip .3cm
\noindent {\it Observatorio Astron\'omico F\'elix Aguilar (OAFA\footnote{
F\'elix Aguilar Astronomical Observatory
(http://www.oafa.fcefn.unsj-cuim.edu.ar/).}, San Juan)}

The F\'elix Aguilar Astronomical Observatory (OAFA), now under the
Faculty of Exact, Physical and Natural Sciences of the National University
of San Juan (UNSJ), was inaugurated on September 28th, 1953 by a small
group of professors from the Engineering School (nowadays a Faculty) of the
National University of Cuyo, led by Carlos Cesco, with the aim of forming
geographers. Its main astronomical facilities are in the Andes Range in
the province of San Juan, in the locality of Calingasta, named Estaci\'on
de Altura (Altitude Station) Carlos Cesco. Over the years, and thanks to
international cooperation, the OAFA has consolidated a strong profile in
the field of Positional Astronomy, which still distinguishes it nowadays.

Fundamental in the history of the institution has been its agreement
with the University of Yale. Between 1973 and 1974, as the University of
Columbia had withdrawn from its agreement with Yale, an agreement was
signed between the newly created National University of San Juan and the
University of Yale. This Yale Southern Observatory-OAFA agreement is
still current today, and its main scientific objective is linked to the study of
Proper Motion of more than thirty million stars. At present, there is also
an agreement with the Chinese Academy of Sciences to operate a Satellite Laser 
Ranging station. Other current activities in the OAFA focus on
Planetary Systems and the study of the sun, activities which are mostly
performed by engineers.

\vskip .3cm
\noindent {\it Instituto Argentino de Radioastronom\'{\i}a (IAR\footnote{
Argentine Institute of Radioastronomy
(http://www.iar.unlp.edu.ar/).}, La Plata)}

The Argentine Institute of Radio Astronomy (IAR) was created in 1962,
only 22 years after the publishing of the first radio astronomical observation,
which proves the highly pioneering character of Argentinian Astronomy in
the mid-twentieth century.

The idea behind IAR first emerged as a project for the creation of a
regional Radio Astronomy Observatory in South America, as an initiative
of Merle Tuve of the Carnegie Institution of Washington. The project began
with the installation of a 30 meter antenna, with the active participation
of Carlos Varsavsky. Later, a second antenna of similar characteristics was
assembled, allowing important surveys of neutral hydrogen to be carried
out. Neither antenna is operating nowadays, although there are plans to
have one of them running again as a School Telescope. As is the case with
other institutions in Argentina, the IAR has suffered from the political and
economic turbulence of the country, affecting its initial growth.

%------------------------------------ 9

As from 2002, the IAR started to perform assignments related to technology 
transfer for the National Commission of Space Activities. Nowadays,
the IAR is strongly committed to technology transfer and continues to perform 
astronomical research, for which it has an active scientific staff of ten
researchers in a variety of fields, being particularly productive in the area
of High Energy Astrophysics.

\vskip .3cm
\noindent {\it Instituto de Astronom\'{\i}a y F\'{\i}sica del Espacio (IAFE\footnote{
Institute of Space Astronomy and Physics
(http://www.iafe.uba.ar/).}, Buenos Aires)}

The Institute of Space Astronomy and Physics (IAFE) is under CONICET and was 
created in 1969 as a result of a restructuring of the {\it Centro
Nacional de Radiaci\'on C\'osmica} (CNRC, National Centre for Cosmic Radiation). 
The institute included members from the CNRC, the Faculty of
Exact and Natural Sciences, and astronomers from the OALP, among which
was the first director of IAFE, Jorge Sahade. The IAFE emerged as a pioneer 
institute in experiments in cosmic radiation, in $\gamma$, X and infra-red rays,
with a strong observational profile, taking great advantage of technological
developments carried out in the country.

At present, the IAFE has more than seventy researchers in different
astronomical subjects, some of which have started in recent years as a result
of the migration of researchers from other national institutions. Due to its
location in the city of Buenos Aires and within the University of Buenos
Aires, which is the biggest in the country, the IAFE has experienced steady
growth.

\vskip .3cm
\noindent {\it Complejo Astron\'omico El Leoncito (CASLEO\footnote{
Leoncito Astronomical Complex
(http://www.casleo.gov.ar/).}, San Juan)}

During the sixties, the OALP purchased a Ritchey-Chr\'etien optical reflector 
telescope, with a 215cm mirror. The installation carried out in cooperation with 
the Universities of C\'ordoba and San Juan, together with
CONICET, as a national observation facility. After a process of searching
and characterization of potential astronomical sites, an area was chosen
close to the OAFA altitude station, in the province of San Juan.

%------------------------------------ 10

For budgetary problems, it was not possible to set up the telescope
at the site selected, and it was installed nearby, but at a lower altitude,
with astronomical conditions inferior to those of the first site. Nowadays,
the agreement between the universities and CONICET in relation to the
CASLEO has expired, and a new agreement is being negotiated.

Despite having started as a services institution, over the years CASLEO
acquired a small astronomical staff. However, with the creation of the Institute 
of Earth and Space Astronomical Sciences (ICATE) in 2009, these
astronomers became part of the new institution, and so CASLEO has returned to 
its original format, with a staff mainly consisting of engineers
and technicians.

Although this telescope has lost its competitiveness, due mainly to its
age, it is still in high demand by the astronomy community in Argentina, especially 
by stellar astronomers. Numerous PhD theses have been produced
in the last twenty years with data obtained through this telescope, and
it is still the astronomical instrument most widely used by the astronomy
community in Argentina.

Besides the 215cm Jorge Sahade telescope (named after the outstanding
Argentinian astronomer), CASLEO also has other instruments: the Helen
Sawyer Hogg 60cm telescope, in Cerro Burek (the site originally chosen for
the 215); the Solar Sub-millimeter Telescope, built by international collaboration 
with the Mackenzie Centre for Radio Astronomy and Astrophysics
of Brazil; and finally, the Astrograph for the Southern Hemisphere, also
installed in Cerro Burek in collaboration with the Astrophysical Institute
of Andaluc\'{\i}a Spain.

\vskip .3cm
\noindent {\it Instituto de Ciencias Astron\'omicas, de la Tierra y del Espacio 
(ICATE\footnote{Institute of Earth and Space Astronomical Sciences
(http://www.icate-conicet.gob.ar/).}, San Juan)}

Founded in 2009, the Institute of Earth and Space Astronomical Sciences
(ICATE) is the youngest astronomical institute in the country, and its
members are mostly former scientific staff from CASLEO. Nowadays it has
about ten researchers, who carry out studies in areas such as Galaxies, Solar
Physics, Stellar Physics and Cosmochemistry. Some members of ICATE are
also professors at the Faculty of Exact, Physical and Natural Sciences at
the University of San Juan, teaching in the {\it Licenciatura}\footnote{
Licenciatura is a five-year first degree obtained in Argentinian Universities.
The following degree would be a PhD.} and PhD in Astronomy.

Among other institutions where astronomical research takes place are
the Institute of Physics of Rosario (Rosario), the Institute of Geological
Correlation (Tucum\'an), the Regional Centre of Scientific Research (Mendoza) 
and the Institute of Detection Technologies and Astroparticle Physics
(Buenos Aires and Mendoza).

%--------------------------------------- 11

\section {Observational Facilities}

\vskip .3cm
\noindent {\it Gemini}\footnote{http://www.geminiargentina.mincyt.gob.ar/
}

Since its beginnings, Argentina has participated in the International
Gemini Partnership, which operates two eight meter diameter optical/infrared 
telescopes, one in Chile and the other in Hawaii, USA. Despite
having a small share in the partnership (approx. 2.5\%), this infrastructure
is highly regarded by the Argentinian astronomy community, since it offers
a concrete possibility to access the latest generation of telescope.

The participation of Argentina in this project has not always been constant or easy. 
In the past, two factors constituted an obstacle to the normal
development of the project. Firstly, the project arrived here as an initiative
of the United States, instead of being suggested by the local community.
Secondly, during several years the project did not have decisive support
from the national institutions responsible for its financing and management.
As a consequence, there was not only scant participation of Argentina in
the political, technical and scientific discussions regarding the project, but
also the loss of actual telescope time between the years 2003 and 2008.

In 2009, the {\it Ministerio de Ciencia, Tecnolog\'{\i}a e Innovaci\'on 
Productiva} (MinCyT\footnote{Ministry of Science, Technology and Productive Innovation
(http://www.mincyt.gov.ar/).
}) took over the project; this, together with the commitment
of numerous astronomers, led to the gradual normalization of Argentinian
participation in the partnership. Nowadays, Argentina has its payment contributions 
up to date, has adhered to the renovation of the international
partnership, and it actively participates in the different international committees.

Argentina is eligible for a total of forty hours observation-time per
semester in both telescopes. On average, the demand is fifty percent higher
than the time actually available.

\vskip .3cm
\noindent {\it CASLEO}

As mentioned above, CASLEO includes a series of operating instruments. 
For most of these, telescope time allocation depends on a scientific
committee made up of representatives from the three Universities involved
and CONICET. Time is assigned each semester, and in most cases the
demands for observation time are satisfied.

%---------------------------------------- 12

The Jorge Sahade Telescope, with a 215cm diameter primary mirror,
contains a variety of instruments: direct CCD of 1340 × 1300 pixels; two
spectrographs, one of which is \'Echelle type, and two photopolarimeters,
one of which was designed and built in CASLEO.

\vskip .3cm
\noindent {\it High Altitude Station "Carlos U. Cesco"}

The main astronomical facilities of the OAFA are found in the locality
of Barreal, San Juan. Some of the operating instruments are: Photoelectric
Astrolabe; Double Astrograph Telescope (in agreement with Yale Southern
Observatory); Automatic Meridian Circle (in agreement with the Royal
Observatory of the Spanish Navy). There are also two instruments for the
study of the sun: MICA (Mirror Coronograph for Argentina) and HASTA
(H$\alpha$ Solar Telescope for Argentina).

\vskip .3cm
\noindent {\it Bosque Alegre}

The 154cm diameter telescope of the OAC, located in the C\'ordoba hills,
had been out of use for many years. At the end of 2008 it was decided to put
it back into operation as an educational telescope and for scientific uses,
and this project received a major boost during 2011. It is expected that it
will soon be possible to invite proposals for observation.

\vskip .3cm
\noindent {\it Pierre Auger Observatory}\footnote{http://www.auger.org.ar/}

The Pierre Auger Observatory, the result of the collaboration of nineteen
countries including Argentina, is located in the province of Mendoza and
is studying ultra-high energy cosmic rays. Although a correlation has been
suggested between cosmic rays and extragalactic objects, these results have
not been confirmed, and therefore the participation of astronomers in the
field is still limited. The positive results of Argentinian collaboration in this
project are commonly used as an example of the capability of our country
to successfully host scientific projects of great magnitude.

\section {Computational Resources}

In Argentina there are diverse groups with significant experience in the
utilization and undertaking of simulations and of numerical models. Consequently, 
there has been a need to acquire high performance computational
resources. Nowadays, the computational infrastructure available at Argentinian 
institutions is somewhat obsolete, so that they have to recur to the
infrastructure of other countries through international collaborations. Recently, 
the MinCyT created the High Performance National Computation
System, which aims to integrate computer clusters throughout the country
into one distributed calculation facility, which will enable an optimal use of
this technology. This is a tentative project in formation, which, if successful, 
will allow the Argentinian astronomical community to access significant
computational resources.

%---------------------------------------------- 13

In 2009, with the involvement of most Argentinian astronomical institutions, 
the {\it Nuevo Observatorio Virtual Argentino} (NOVA\footnote{New Virtual Argentinian Observatory
(http://www.nova.org.ar/).}) was created,
aimed at the national and international coordination of resources relating to
data-centres, software tools and information about inter-operational standards. 
Recently, the International Virtual Observatory Alliance\footnote{http://www.ivoa.net/} accepted
NOVA's membership; the new virtual institution is now studying the first
projects to be run, as well as searching for sources of funding to permit
sustained medium-term operation.

\section {The Argentinian Astronomical Society}

The {\it Asociaci\'on Argentina de Astronom\'{\i}a} (AAA\footnote{Argentinian Astronomical Society
(http://www.astronomiaargentina.org.ar/).}), which has recently
turned fifty years old, brings together the vast majority of Argentinian
astronomers, with the main objective of the promotion and progress of astronomy 
and its related activities in the country. Its specific functions are:
to provide appropriate spaces for discussion and promotion of research activities 
in astronomy and in related sciences; supply institutional back-up
to initiatives in the area; organize scientific meetings where astronomers
may exchange information; give support to young astronomers by means
of scholarships and work-contracts; act as a contact between astronomers
and organizations which support science in Argentina and the world; and
finally, the outreach of astronomy among the general public.

The AAA is economically sustained by its members' fees, and is administered by a 
Board elected by the member's general Assembly, renewed
every three years. The Assembly also votes for the National Committee of
Astronomy, which is responsible for liaison with the International Astronomical 
Union (IAU\footnote{http://www.iau.org/}).

The AAA also has an important publishing role. The {\it Bolet\'{\i}n de la Asociaci\'on 
Argentina de Astronom\'{\i}a} (BAAA) is produced annually by an editorial committee, 
which is independent of the Board. Following a refereeing process, the BAAA publishes 
the papers presented during the annual
meetings, where numerous astronomers and undergraduates get together,
creating a good environment for scientific and policy discussions. Special
debates or sessions on specific topics are held. The AAA also publishes the
Workshops and Books series, as well as a newsletter. As from 2006, the AAA
holds annual workshops related to a variety of topics. Among the recent
have been Observational Astronomy, Theoretical Astronomy, the outreach
of Astronomy, History of Argentinian Astronomy, the Gemini Project and
Computational Astronomy.

%------------------------------------- 14

The Argentinian Astronomical Association grants three prestigious
awards: the {\it Carlos Varsavsky Award} for the best Doctoral Thesis, the {\it Jos\'e
Luis S\'ersic Award} for the outstanding Senior Researcher and the {\it Jorge
Sahade Award} for Scientific Trajectory. The first two awards are granted
every two years, and the latter, every three.

\section {Education in Argentina}

Argentina has historically had a very high literacy level among its population, 
and a great level of achievement of its University graduates. Nevertheless, 
there are nowadays some warning signs, especially as regards
Elementary and Middle education levels. Clearly reflecting the problem is
the noticeably low achievement of Argentinian secondary students in the
last PISA ({\it Programme for International Student Assessment}\footnote{http://www.pisa.oecd.org/
}), an exam
for 15-year old middle school students, which located Argentina in the 58th
place. Secondary school education, of course, affects studies at university,
which has led to the creation of preparatory courses in order to bring students' 
knowledge and training to the level of the new demands they need
to face at university.

Traditionally, the educational system in Argentina has been public
and free, including universities. Moreover, the highest academic levels and
achievements are well known to have originated and been found in public universities, 
despite the rapid growth in the private education sector
nowadays. Free education in Argentina at all levels has permitted unlimited access to 
higher education for all social levels, with a strong presence
of the middle class.

The vast majority of Argentinian astronomers studied for their degrees
in Argentina, and a large number have undertaken postdoctoral studies
abroad. Most researchers who nowadays carry out astronomy studies in our
country were formed and trained in the Universities of La Plata, C\'ordoba
and San Juan, although some have graduated in Physics.

In Argentina, about ten students per year obtain their PhD with an
astronomical orientation, in one of the following institutions:

\vskip .3cm
\noindent {\it Facultad de Ciencias Astron\'omicas y Geof\'{\i}sicas de La Plata 
(FCAGLP\footnote{Faculty of Astronomical and Geophysical Sciences}),
La Plata}

%-------------------------------------- 15

As mentioned above, this was the first School of Astronomy in the country. 
Here students work towards the Licenciatura and PhD in Astronomy,
the latter having the highest category given by the Comisi\'on Nacional de
Evaluaci\'on y Acreditaci\'on Universitaria (CONEAU\footnote{National Commission 
of University Evaluations and Accreditations (http://www.coneau.edu.ar/).}). 
Although the average duration 
of a Licenciatura is of five years (containing 30-35 subjects),
it is taking students longer lately.

\vskip .3cm
\noindent {\it Facultad de Matem\'atica, Astronom\'{\i}a y F\'{\i}sica 
(FaMAF\footnote{Faculty of Mathematics, Astronomy and Physics.}), C\'ordoba}

This is the second School of Astronomy founded in the country, and
this and the one in La Plata are the two most traditional ones. Although
FaMAF is independent of the Astronomical Observatory, the professors
of the Observatory are in charge of the academic formation of astronomy
students in these specific subject-areas. FaMAF students work towards the
Licenciatura and PhD in Astronomy, and the latter has the highest category
given by the CONEAU. As in La Plata, the average period of studies for
the Licenciatura has extended, possibly due to its demanding level, which
may be compared to a Master studies in Universities abroad.

\vskip .3cm
\noindent {\it Faculty of Exact, Physical and Natural Sciences of the 
UNSJ\footnote{http://www.unsj.edu.ar/facu exacta.php}, San Juan}

Since 1995, students can work for the Licenciatura in astronomy in San
Juan. This was originally a four-year degree, but nowadays it has been
extended to the same period as in other Universities. Not long ago, there
was no possibility of completing a PhD in astronomy or in related sciences
in San Juan; consequently, most graduates then migrated to other national
or international universities, to complete their postgraduate studies. Some
of these researchers are now returning to San Juan, presaging a powerful
growth for local astronomy. In 2009, a PhD program in Astronomy was
started, specializing in Observational Astronomy.

An important contribution to research in the area of Astronomy has
come from doctors in physics, mainly from the University of Buenos Aires
(UBA\footnote{http://www.uba.ar/}). Thanks to its strategic location and joint dependency on the UBA
and CONICET, IAFE attracts physics students from the Faculty of Exact,
Physical and Natural Sciences, who are interested in astronomical topics.

%-------------------------------------------- 16

\section {Institutions and Resources for the Promotion of Science}

According to the 2010 UNESCO Science 
Report\footnote{
http://www.unesco.org/new/en/natural-sciences/science-technology/prospective-studies/unesco-science-report/unesco-science-report-2010/}, Argentina has the third
largest gross domestic product (GDP) in Latin America, after Brazil and
Mexico (2009 figures), while its Gross Expenditure on Research and Development 
(GERD) represents 0.51\% of the GDP, setting Argentina again in
the third place, after Brazil and Chile (2007 figures). Compared to other
countries of the region, Argentina has a low percentage of funds provided
by the business sector, while a high percentage of the scientific activity
carried out in Argentina is financed by the State.

In 2007, the Argentine government transformed the previous Secretary
of Science and Technology into the Ministry of Science, Technology and
Productive Innovation (MinCyT), with CONICET as one of its most important 
agencies for the development of science. The national universities are
the other pillar sustaining the Argentinian Science System. Other institutions 
which provide resources for the promotion of science are the National
Agency for the Promotion of Science (also dependent on the MinCyT), as
well Secretariats and Ministries dependent on the provinces.

Perhaps as a result of the constant crises the Argentinian scientific system has 
been subject to, its astronomical community has acquired great
experience in obtaining funds abroad, in many cases as a result of international 
scientific collaboration programs. Contributions from private national funds in 
Argentina are hardly existent. Nonetheless, a great part of
the Argentinian scientific community remembers with gratitude the Antorchas Foundation, 
which before disappearing in 2006, financed numerous
astronomical projects.

\vskip .3cm
\noindent {\it CONICET}

CONICET is, no doubt, one of the main driving forces behind scientific
development in Argentina. Since its creation more than half a century ago,
CONICET has implemented the so-called "career of scientific researcher"
(Carrera del Investigador Cient\'{\i}fico y Tecnol\'ogico, CIC), by which it hires
researchers to enable them to carry out their activities in one of the national
scientific institutions. Nowadays, CONICET has 6350 researchers, of which
114 are in the area of astronomy (2010 figures).

The CIC has a scale of five categories, access to which includes peer
evaluation; except for the first category (assistant researcher), permanency
in the career is guaranteed by merely achieving a passing mark in the biannual 
evaluations. Following long decades of scarce income and a lack of
senior researchers, a radical change in the last decade has provided CONICET with 
approximately 500 annual positions for all areas of research,
which allows, within the range of the exact sciences at least, inclusion in
the CIC of a large proportion of those who qualify for it.

%------------------------------------- 17

Currently, around ten astronomers qualify to enter the CIC annually,
which has resulted in the sustained growth of Argentinian astronomy in all
its aspects. The qualification criteria to enter the CIC are basically having
graduated as a PhD and having published internationally refereed articles
in first level journals. However, one drawback of this policy is the limited
mobility of researchers who, in many cases, obtain a permanent position
without undertaking previous postdoctoral activities. CONICET, aware of
this complication, has implemented scholarships for researchers who are
interested in working abroad, though this measure has not yet been shown
to be successful.

Another important input of CONICET in the human resources area
is the award of doctoral scholarships. Most doctoral theses in astronomy
undertaken in Argentina are funded by CONICET scholarships, with a duration 
of five years. CONICET also contributes funds for research projects
which are quite accessible for most researchers. Nevertheless, these funds
only cover the basic expenses of a group of researchers and part of their
travel expenses, thus limiting their presence and participation in international 
scientific events or scientific visits to other research centres. The
typical ratio between funds in research grants and salaries ranges between
0.1 and 0.3, significantly lower than international standards.
CONICET promotes research by creating and financing scientific institutions. 
At present, most researchers in the area of astronomy work in one
of the following institutions: IAFE, IALP, IAR, IATE, ICATE, CASLEO,
IFIR\footnote{Instituto de F\'{\i}sica de Rosario
(http://www.new.ifir-conicet.gov.ar/).}, CRICYT\footnote{Centro Regional de 
Investigaciones Cient\'{\i}ficas y Tecnol\'ogicas (http://www.mendoza-conicet.gob.ar/).}, 
INSUGEO\footnote{Instituto Superior de Correlaci\'on Geol\'ogica
(http://www.insugeo.org.ar/).}, ITEDA\footnote{Instituto de Tecnolog\'{\i}a en 
Detecci\'on y Astropart\'{\i}culas (http://www.iteda.org/).}. The present policy of CONICET is 
that the institutes will ideally work in co-management with the
National Universities, as is the case of several of the institutions mentioned
above. CONICET also provides funding for organizing scientific meetings
and, particularly as regards astronomy, for the Argentinian membership of
the IAU and for the {\it Astronomy \& Astrophysics} journal.

\vskip .3cm
\noindent {\it National Universities}

Universities are undoubtedly the principal actor in Argentinian scientific
development. They are responsible for undergraduate and graduate studies,
without which there would be no science in Argentina. It is important to
remember and point out that the public university is free of charge in
Argentina and in some cases, low-income students may get some financial
help.

%----------------------------------------- 18

The universities in Argentina are autonomous and so it is difficult to
generalise about them. However, the most relevant aspects of all the universities 
are as follows: universities provide an important number of jobs for
researchers in astronomy; these may be full-time ("exclusive") or part-time
("semi" or "simple") posts; in the first two categories, undertaking scientific
research is compulsory. At present, very few new posts are being created,
and thus the possibilities for entering the university system are mainly limited 
to vacancies. Usually, researchers may qualify for these posts through
a selection committee, and this process may be renewed periodically.

Universities also finance doctoral scholarships and provide grants for
basic expenses of research projects. It is important to point out that universities 
support the running of Observatories and co-finance CONICET
institutes with double dependency. In fact, a large number of professors are
also CONICET researchers.

\vskip .3cm
\noindent {\it Agencia Nacional de Promoci\'on Cient\'{\i}fica y Tecnol\'ogica 
(ANPCT\footnote{National Agency for the Promotion of Science and Technology.})}

Dependent on the MinCyT, the ANPCT has as one of its central goals
the funding of large-scale research projects; however, in our case, these are
hard to obtain due to the fact that the area of astronomy is not considered
a priority.

\section {Scientific Meetings}

The Argentinian astronomical community has a long tradition of holding
scientific events. The most traditional one is the annual meeting of the
AAA, which brings together more than 300 students and graduates in a
different part of the country every year. For the 2011 series, and for the
first time, the event was jointly organized with the Chilean Society of Astronomy.

Numerous international scientific meetings have taken place in Argentina in the 
last fifty years. In 1991, Buenos Aires hosted the 21st IAU
General Assembly; in the years 1983 and 2001, two IAU Latin American regional 
meetings were held, the third in Buenos Aires and the tenth
in C\'ordoba, respectively. Since 1969, eight IAU Symposiums have taken
place: {\it The Problem of the Variation of the Geographical Coordinates in the
Southern Hemisphere} (IAUC 1, La Plata 1968); {\it Spectral Classification and
Multicolour Photometry} (IAUS 50, C\'ordoba 1971); {\it Wolf-Rayet and High-Temperatures 
Stars} (IAUS 49, Buenos Aires 1971); {\it Solar Gamma-, X-, and
EUV Radiation} (IAUS 68, Buenos Aires 1974); {\it Evolutionary Processes in
Interacting Binary Stars} (IAUS 151, C\'ordoba 1991); {\it Eruptive Solar Flares}
(IAUC 133, Iguaz\'u, 1991); {\it Jets at All Scales} (IAUS 275, Buenos Aires
2010); and {\it Comparative Magnetic Minima: Characterizing quiet times in
the Sun and Stars} (IAUS 286, Mendoza 2011). There is currently a special
emphasis on the organization of scientific meetings: during 2011, at least
nine events of different features and topics were organized.

%----------------- 19

\section {Promoting Astronomy within the Society at Large}

The outreach of astronomy to the general public in Argentina has grown
significantly in the last few years. Historically, it was an activity reserved for
only a few and, even though the traditional institutions carried out related
activities, these were not innovative and were mainly related to specific
astronomical events or to the systematic attention to observatory visitors.

Nowadays, a great part of the astronomical community is seen to be
clearly committed to performing extension activities for the general public.
No doubt, the International Year of Astronomy in 2009 gave a mighty
impulse to the development of these activities, which were not limited to
traditional institutions such as the OLP or OAC, but expanded to the
vast majority of astronomical institutions, as well as to others not devoted
directly to scientific investigation, but which also took on a great role in
the promotion of astronomy.

OAC, OALP, and OAFA have their own museums, while there are two
traditional planetariums, in Rosario and, the most emblematic one, the
Galileo Galilei Planetarium in the city of Buenos Aires, which has recently
been refunctionalized to include full-dome digital technology. In the city
of Malarge, in the province of Mendoza, there is a high technology digital
planetarium; in San Luis, the University of La Punta also operates one.
Moreover, the National University of La Plata is building a large digital
planetarium, while in C\'ordoba an opto-mechanical planetarium is being
built, based on an instrument donated by the city of Nantes in France, and
the possible construction of a digital one is being studied. Three examples 
of astronomical promotion at non-traditional sites are the La Punta
Astronomy Park in the city of San Luis\footnote{http://www.palp.edu.ar/}, 
the Sky Plaza in the city of
Esquel\footnote{http://www.plaza-del-cielo.com.ar/} and the awi Puna Observatory 
in the small locality of Tolar Grande
(150 inhabitants), Salta, 3500m above sea level in the middle of the Andean
Puna.

%--------------------------- 20

\section{Amateur Astronomy}

The fascination astronomy has always evoked has not escaped the Argentinian 
people, and that is why there are numerous Friends of Astronomy
associations, reflecting a long tradition of interest in astronomy in the country. 
The most important of these groups is the Argentinian Association of
Friends of Astronomy (AAAA\footnote{http://www.asaramas.com/}), 
which dates back 81 years and is thus
older than the professional association AAA, and is the oldest of its kind
in Latin-America.

On some occasions, amateur astronomers also engage in scientific
projects, for example devoted to precision astrometry of minor bodies of
the Solar System. The Association of Argentinian Observatories of Minor
Bodies\footnote{http://www.aoacm.com.ar/} groups the amateur observatories accredited 
by the Minor Planet Centre\footnote{http://www.minorplanetcenter.net/iau/mpc.html}. 
At present, there are fourteen amateur observatories which hold
a certification of the Minor Planet Center of the IAU.

\section{Future Astronomical Projects}

Except for Gemini, it is noticeable how far behind Argentina is in terms
of latest generation observational infrastructure. For this reason, the Argentinian 
astronomical community has an agenda of numerous new astronomical infrastructure projects, 
some of which are national projects, while
others are international collaborations. Recently, the National Ministry of
Science and Technology has created a committee of Astronomy and Sciences of the Universe, 
with the purpose of evaluating projects in these
areas. After inviting the scientific community to present possible projects,
and a prior evaluation process, the committee has selected and ranked four
greatly varied projects for possible funding.

The projects are the following:

\vskip .3cm
\noindent {\it Long Latin American Millimeter Array (LLAMA)}

This is a bi-national project in collaboration with Brazil, for the installation of 
one or two antennas in the millimeter/sub-millimeter wavelengths
in the northwest of Argentina at a height of 4600m. The project includes
the possibility of setting up a Very Long Baseline Interferometer by means
of integration with projects such as ALMA\footnote{http://www.almaobservatory.org}, 
ASTE\footnote{http://www.ioa.s.u-tokyo.ac.jp/∼kkohno/ASTE} and/or 
APEX\footnote{http://www.apex-telescope.org} in
the Chilean northeast. The Argentinian side is led by IAR researchers.

%--------------- 21

\vskip .3cm
\noindent {\it Cherenkov Telescope Array (CTA)}

This project consists of a mega-international collaboration to build and
operate the next generation of ground-based very high energy gamma-ray
instruments to provide a deep insight into the non-thermal high-energy
universe. CTA is made up of 25 countries and is considering the building of
different types of telescopes in both hemispheres. Some of the Argentinian
institutions involved in the project are IAFE, IAR and ITEDA. One of
the main reasons for Argentina to participate in this project is its possible
location within Argentinian territory, probably in the northwest or at sites
close to CASLEO in the province of San Juan.

\vskip .3cm
\noindent {\it Doubling the Time at Gemini}

This project consists in doubling Argentina's time in the international
Gemini partnership, which would be possible due to the withdrawal of Great
Britain. This proposal is not led by any particular institution, but has the
support of numerous astronomers and institutions at national level.

\vskip .3cm
\noindent {\it Argentina-Brazil Astronomical Centre (ABRAS)}

This is a bi-national project which consists in the joint installation of a
1m/1.2m infrared robotic telescope with a medium size field of view in the
Macon site, Salta Province at 4600m above sea level where an 8m-diameter
dome is under construction. The Argentinian side is led by the IATE.

\section{Human Resources}

As mentioned, the national universities and CONICET are responsible for
almost all astronomical posts available in Argentina. If we take into account
teaching and research positions, the number amounts to approximately 250
scientists (considering that many of them depend on both these institutions). 
Fig. 5 shows the number of CONICET researchers by category
(2010 figures). If we assume that, from the category of "Independent" and
upwards, we are looking at senior researchers, we may conclude that the
population pyramid is reasonable, which has been achieved during recent
years thanks to the steady entry of young researchers.

Most PhD students undertake their studies with some kind of scholarship, the 
most numerous being those of CONICET. There is also a possibility of access 
to university scholarships, or ANPC grants. The present
number of postgraduate students and postdoctoral fellows from CONICET
in the area of astronomy is 62, which indicates that the human resources
of astronomy will continue growing in Argentina.

Nowadays, approximately half the astronomical community in Argentina
are women, and these are expected to become the majority in the medium
term, taking into account that women are already a majority among undergraduate 
students. It is important to point out as well that Argentina
has the highest percentage of women members in the 
IAU\footnote{http://www.iau.org/administration/membership/individual/distribution/}, 
48 out of 131.

%--------------- 22

%--------------- Table 2
\begin{figure}[h!]
  \caption{Number of CONICET researchers per category.}
  \centering
    \includegraphics[width=.35\textwidth]{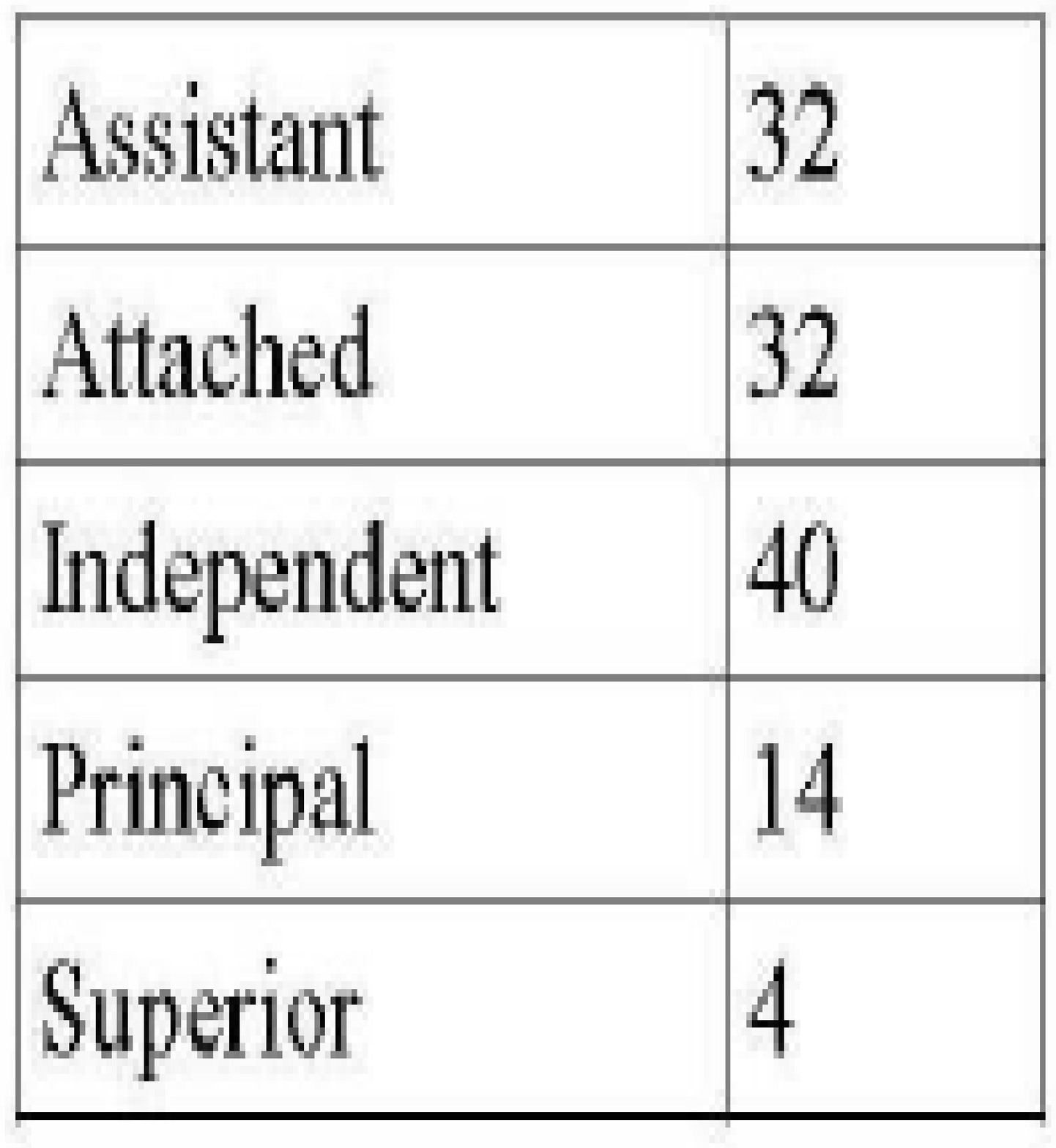}
\end{figure}

Although Argentinian astronomy was initiated by citizens of other countries, with 
Benjamin Gould being NAO's first director, and that this tradition lasted many years, 
nowadays almost all scientific posts are held by
Argentinian citizens. One of the reasons for this is the policies of the institutions 
for the promotion of science, which do not have easy procedures for
the incorporation of foreign citizens into the system; it is currently not possible 
to enter the CIC of the CONICET without Argentine residence, while
the entry of foreign citizens to universities is by means of exception. In the
case of postgraduate student scholarships, there is a CONICET program
aimed specifically for Latin American citizens. Although most postdoctoral
fellows are Argentinian as well, there are programs suited for foreigners.

The history of salaries in the scientific sector in Argentina is complex,
with extreme ups and downs. Nowadays the situation may be qualified as
fairly good, although there is a noticeable difference with other countries
of the region, Brazil or Chile for instance, where scientific activity is much
better paid.

Due to successive economical and political crises, as well as to the search
for new scientific horizons, a great number of Argentinian astronomers reside in 
foreign countries, many of whom have gained international prestige.
Nowadays, most astronomers migrate to attain training and expertise, and
with the hope of coming home after completing their postdoctoral activities. It is 
important to point out that many Argentinian astronomers living
abroad maintain ties with their local peers, and that this is supported by
national subsidies devoted specifically to this kind of interchange, as in
the case of the PICT-Ra\'{\i}ces (Roots) grant of the ANPCT and the Cesar
Milstein\footnote{http://www.raices.mincyt.gov.ar/aplicar\_milstein.htm} 
grant of the MinCyT.

%----------- 23

\section{Scientific Production}

The Digital Library for Physics and Astronomy for the Astrophysics Data
System (ADS\footnote{http://www.adsabs.harvard.edu/}) shows that about 
300 articles\footnote{This number should be taken only as an approximate figure due to the possible
incompleteness of ADS as well as the techniques used to choose the articles, which may
include thematic contaminations or omissions.} with at least one author
from an Argentinian institution were published during 2010. If we assume
that the number of active astronomers, including postgraduate students,
is approximately 250, the result is a little more than one article for each
researcher, similar to all other sciences in Argentina. If we consider only the
four non-specialist publications used, {\it The Astrophysical Journal (ApJ)}, 
{\it The Astronomical Journal (AJ)}, {\it The Monthly Notices of the Royal Astronomical
Society (MNRAS)}, and {\it Astronomy \& Astrophysics (A\&A)}, the total of
publications in the same period is 105, distributed as follows: A\&A (38),
MNRAS (35), ApJ (22) and AJ (9), which shows a clear preference of the
Argentinian astronomers for publishing in European journals.

This may be explained principally by the high costs per page in North
American publications, which are hard to afford with the limited grants
given in Argentina. Needless to say, Argentinian astronomers do not pay
for publication in A\&A, due to Argentinian membership in the A\&A partnership since 2004; 
actually, Argentina was the first non-European country
to be incorporated to the A\&A.

\section{Technological Development}

From its beginnings, Argentinian astronomy brought about and has been
associated with promising instrumental development. It is sufficient to remember Charles 
Perrine, who built a 76cm reflecting telescope in the OAC
at the beginning of the twentieth century. Nonetheless, with the passing
of the years, this activity has lost competitiveness and support. There is a
general consensus that this situation should be reversed, but to date there
is no certainty regarding the possibility of making it happen in the short
term. One of the paths being explored is to encourage high technology companies, 
such as INVAP\footnote{http://www.invap.com.ar/}, to participate, in conjunction with astronomers
and astronomical institutions, in the development of future instruments for
Gemini.

The IAR is also an institution of great potential; although the advanced
technological development carried out there is mainly aimed at the satellite
area, the necessary tools and expertise exist there to reorient it to the
development of radio-astronomical instruments.

%--------------- 24

\section{Argentina as a Site for Large Astronomical Facilities}

The western border of Argentina lies on the high summits of the Andes
Range, which extends over 4000km. Moreover, between the central region
and the northern border there are significant peaks and great plains at
considerable altitude, such as the Puna which, 150km wide and 350km
long, and an average of about 4000m in altitude, constitutes one of the
major highlands in the world. The climatic characteristics of this region
are of a very dry area with a high percentage of clear nights, similar to
those in the north of Chile.

All these features of the region have generated numerous initiatives from
the astronomical community to attract important international projects to
Argentina. Since the year 2000 there has been intense activity, aimed not
only at selecting potential sites, but also advancing in their characterization.
Probably the work with the greatest breadth in this direction has been undertaken 
by researchers of the IATE, who were in charge of the astronomical
characterization of the Cerro Mac\'on, in the Province of Salta. This study
was carried out in agreement with the European Southern Observatory\footnote{http://www.eso.org/}
within the framework of site selection for the European Extremely Large
Telescope\footnote{http://www.eso.org/sci/facilities/eelt}. Though Argentina was 
not finally chosen for the installation
of the telescope, the region has been thoroughly characterized in all its
aspects, which has given rise to new projects such as ABRAS, mentioned
above. Under the IAR's responsibility, this site was studied in reference
to its vapour content, as well as other regions of the Puna, such as Alto
Chorrillo, which is being considered for the LLAMA project. The region
has also been studied for high-energy projects. The central region of the
Andes, especially the zone close to CASLEO, has also been proposed for
international projects, particularly the Square Kilometre 
Array\footnote{http://www.skatelescope.org/}. Led by
researchers of the IAFE and the IAR, Argentina actively participated to
establish this project in the region. Although Argentina was not selected
for this project either, the Argentinian astronomical community has proved
its strong vocation to encourage such initiatives. What is more, it has been
demonstrated that there are sites of great potential, and thus it is only a
matter of time before a project of great magnitude, as in the case of the
Pierre Auger Cosmic Rays Observatory, is attracted.

%--------------------- 25

\section{Future challenges}

Astronomy in Argentina may be considered as one of the foundational areas
of science in our country. It has a long tradition as well as significant regional
and international recognition. Despite repeated periods of crisis, nowadays
it is experiencing sustained growth, especially in terms of human resources.
Astronomical research is carried out in traditional institutions as well as
in novel ones, in at least ten scientific centres. As long as no more local
or international crises interfere, this tendency of constant growth should
consolidate in the forthcoming years.

Yet the Argentinian astronomical community, together with the institutions 
for the promotion of science, have a series of challenges ahead of them:
to update the observation infrastructure, improve young researcher's mobility, 
improve scientific production and its impact, recover high-technology
development of astronomical instruments, raise the amounts of grants destined to 
research, encourage astronomical research in more regions of the
country, enable greater inclusion of foreign researchers and, last but not
least, attract international astronomical projects of great magnitude to be
set up in the country.

\vskip 1.cm
\noindent {\bf Acknowledgments}
\vskip .1cm
This work would have not been possible without the collaboration of numerous 
colleagues who contributed important data and precision regarding
a diversity of topics. I would like to thank Mario Abadi for his contributions 
and advice, and Galit Shani and Joss Heywood for their help with
the manuscript.

I am also grateful to anonymous referees for their suggestions and linguistic 
improvements brought to the paper.

\vskip .3cm
\noindent {\bf Bibliography}
\vskip .1cm

One of the main sources used for the historical institutional aspects was
the book Historia de la Astronom\'{\i}a Argentina (History of Argentinian Astronomy), 
AAABS 2 from the Book Series of the Argentinian Association
of Astronomy. The other important sources of information were the web
pages quoted in this article, as well as colleagues whom I consulted.

\vskip .3cm
\noindent {\bf Abbreviations}
\vskip .1cm

\noindent AAA: The Argentinian Astronomical Society

\noindent ANPCT: The National Agency for the Promotion of Science and Technology

%-------------- 26

\noindent BAAA: Journal of the Argentinian Astronomical Society

\noindent CASLEO: The Leoncito Astronomical Complex
 
\noindent CIC: Career of the scientific researcher of the CONICET

\noindent CONEAU: The National Commission of University Evaluations
and Accreditations (Comisi\'on Nacional de Evaluaciones
y Acreditaciones Universitarias)

\noindent CONICET: The National Council for Scientific and Technical Research
(Consejo Nacional de Investigaciones Cient\'{\i}ficas y T\'ecnicas)

\noindent CRICYT: Regional Centre for Scientific and Technical Research
(Centro Regional de Investigaciones Cient\'{\i}ficas y T\'ecnicas)
 
\noindent FaMAF: Faculty of Mathematics, Astronomy and Physics

\noindent FCGALP: Faculty of Astronomical Sciences and Geophysics

\noindent IAFE: The Institute of Space Astronomy and Physics

\noindent IALP: Institute of Astrophysics of La Plata

\noindent IAR: Argentinian Institute of Radioastronomy

\noindent IATE: Institute of Theoretical and Experimental Astronomy

\noindent IAU: International Astronomical Union 

\noindent ICATE: Institute of Earth and Space Astronomical Sciences

\noindent IFIR: Institute of Physics of Rosario

\noindent INSUGEO: Superior Institute of Geological Correlation

\noindent ITEDA: Institute of Technologies in Detection and Astroparticles

\noindent MinCyT: Ministry of Science, Technology and Productive Innovation

\noindent NOVA: New Virtual Observatory

\noindent OAC: Astronomical Observatory of C\'ordoba

\noindent OAFA: F\'elix Aguilar Astronomical Observatory

\noindent OALP: Astronomical Observatory of La Plata (nowadays the Faculty
of Astronomical Sciences and Geophysics)

\noindent NAO: The National Argentinian Observatory (nowadays OAC)

\noindent UBA: University of Buenos Aires

\noindent UNC: National University of C\'ordoba

\noindent UNSJ: National University of San Juan

\end{document}